\UseRawInputEncoding
\documentclass[pre,10pt,aps,twocolumn,amsmath,amssymb,superscriptaddress]{revtex4-1}
\pdfoutput=1
\usepackage{hyperref}
\usepackage{graphicx}
\usepackage{color}
\usepackage{multirow}
\usepackage{amsmath}
\usepackage{array}
\usepackage{booktabs}
\usepackage{multirow}
\usepackage{tabularx}
\usepackage{cleveref}

\newcommand{\be}{\begin{equation}}
\newcommand{\ee}{\end{equation}}
\newcommand{\bea}{\begin{eqnarray}}
\newcommand{\eea}{\end{eqnarray}}
\def\c#1{~\cite{#1}}
\def\f#1{Fig.~\ref{#1}}

\def\beq{\begin{equation}}
\def\eeq{\end{equation}}

\newcolumntype{Y}{>{\small\raggedleft}X}
\begin{document}

\definecolor{dkgreen}{rgb}{0,0.6,0}
\definecolor{gray}{rgb}{0.5,0.5,0.5}
\definecolor{mauve}{rgb}{0.58,0,0.82}

\title{Machine Learned Phase Transitions in a System of Anisotropic Particles on a Square Lattice}
\author{Karthik Padavala}
\author{Avaneesh Singh}
\author{Joyjit Kundu}
\affiliation{Department of Physics, Indian Institute of Technology Hyderabad, Kandi, Sangareddy, TS 502285}
\email{jkundu@phy.iith.ac.in}

\begin{abstract}
The area of Machine learning (ML) has seen exceptional growth in recent years. Successful implementation of ML methods in various branches of physics has led to new insights. These methods have been shown to classify phases in condensed matter systems. Here we study the classification problem of phases in a system of hard rigid rods on a square lattice around a continuous and a discontinuous phase transition using supervised learning (with prior knowledge about the transition points). On comparing a number of ML models we find that convolutional neural network (CNN) classifies the phases with the highest accuracy when only snapshots are given as inputs. We study how the system size affects the model performance. We compare the performance of CNN in classifying the phases around a continuous and a discontinuous phase transition. Further, we show that one can even beat the accuracy of CNN with simpler models by using {\em physics-guided} features. Lastly, we show that the critical point in this system can be learned without any prior estimate by using only the information of the ordered phase (as training set). Our study reveals the ML techniques that have been successful in studying spin systems can be easily adapted to more complex systems. 
\end{abstract}

\maketitle

\section{\label{sec:one}Introduction}


The area of Machine learning has seen a tremendous growth in the last two decades. Researchers have made great progress in areas like computer vision\c{8718718}, natural language processing\c{8416973}, medical diagnostics\c{debruyne,8572804} using deep learning methods\c{LeCunn,Schmidhuber_2015}. The availability of a large amount of data and higher computing power due to the advancements in hardware have made the growth possible. Machine learning techniques usually perform better with a large amount of data. Data-rich branches of physics like High energy physics and Astronomy thus have employed Machine learning techniques successfully in extracting physical insights from the data\c{BALL_2010,Baldi_2014,albertsson2018machine,Das_Sarma_2019}. In recent years Machine learning has made its way into other branches like Condensed matter physics and  Statistical physics. Both supervised and unsupervised learning methods have been successfully employed to detect phase transitions and classify different phases\c{Arai_2018,broecker2017quantum,Iakovlev_2018,Carrasquilla_2017,Wetzel_2017,PhysRevB.94.195105,bedollamontiel2020machine,PhysRevLett.120.176401,PhysRevB.96.205146,PhysRevX.7.031038,Boattini_2019,Wang_2016}. Neural networks have shown great potential in identifying new exotic phases and phase transitions even in systems where the order parameter cannot be defined explicitly\c{Li2018MachineLI,PhysRevLett.122.210503}.
Generative neural networks like restricted Boltzmann machines and variational autoencoders have been employed to model physical probability distributions and extract features in spin models\c{PhysRevResearch.2.023266,JMLR:v18:17-527}.

Here, we focus on the specific application of machine learning in classifying different phases separated by a phase transition. Usually, in case of a structural phase transition one can construct a suitable order parameter to distinguish between the different phases. Away from the transition point, one may even be able to identify the phases just by looking at the snapshots. However, this distinction gets blurry as one approaches the transition more and more closely (at least for a continuous or a weakly first-order transition). Especially, in the vicinity of a critical point, large fluctuations are expected to interfere. Now the question is whether one can train a statistical classifier to distinguish between such phases just by using the configurations. The classification of the ordered and disordered phases, and the study of the critical properties of Ising model have been extensively investigated in recent times\c{JMLR:v18:17-527,PhysRevResearch.2.023266,Wang_2016,Mehta_2019,doi:10.7566/JPSJ.86.063001}. After the success of Machine learning methods in characterizing simpler models like Ising model, they are now being explored in studying more complex systems, like liquid crystals. In experiments, liquid crystals are usually studied using optical imaging\c{PhysRevLett.110.057801}. Recently Machine learning techniques are applied to classify phases and predict the physical properties of liquid crystals from the experimental data\c{article,PhysRevE.99.013311,minor2019endtoend}. Convolutional neural networks (CNNs) have been shown to successfully classify nematic and isotropic phases in continuum with very high accuracy\c{Sigaki_2020}. In this case the isotropic-nematic transition is first-order in nature where the order parameter changes abruptly in the vicinity of the transition point. 

In this context, we ask how difficult is the classification task near a continuous transition when compared with the same around a first-order transition. We investigate how different machine learning models perform on this classification task and how the system size affects the performance. We further ask if {\em physics-guided} features help in learning the phases. In addition, we examine if the method of {\em Learning by confusion} can be extended to the system of rods to determine the critical (without any prior estimate) using the information about the ordered phase\c{huber2017,Tan_2020}. This technique is quite powerful as the usual supervised ML techniques rely on the prior knowledge of the critical point to be trained-- this may not always be possible for all complex systems\c{huber2017}. Understanding these issues can help us extending ML techniques to more complex systems and also tackling systems where the identification of structural order (if exists) is nontrivial, e.g., amorphous systems\c{deepmind} .

To investigate the above questions, we study a model of long hard rods on a two-dimensional square lattice that undergoes isotropic-nematic-disordered phase transitions with increasing density. Note that, systems of hard rectangles with finite width exhibit more complex liquid crystalline phases\c{PhysRevE.89.052124}. Such hard core lattice gas models are, in general, relevant for understanding self-assembly of nanoparticles\c{phil2003}, glass transition\c{glass_lat}, adsorption of gas molecules on metal substrates\c{Phys.Rev.B32.4653,PRL54.1539,binder2000} and entropy-driven phase transitions (realized in liquid crystalline assemblies of various colloidal systems\c{kuij2011,kuij2012,Lekkerkerker1996,Zhao2011,dogic_pnas}). In contrast to the continuum model, the high-density phase of the system of hard rods on a lattice is an orientationally disordered phase that remained inaccessible until recently due to large relaxation times\c{kundu2,PhysRevE.87.032103,PhysRevE.89.052124}. Using a novel Monte Carlo algorithm with nonlocal moves one can thermalize the system at very high densities to show that it undergoes two continuous transitions with increasing density: first from a low-density isotropic to an intermediate density nematic phase and second, from the nematic phase to a high-density disordered phase\c{PhysRevE.87.032103,PhysRevE.89.052124}. We generate Monte Carlo sampled data for the hard rods system on a square lattice to classify the isotropic (I) and nematic (N) phases around  the first I-N critical point using various Machine learning techniques. We show how the system size affects the performance of the models. Note that this classification task is trivial when the I and N phases are far from the critical point. However, closer to the critical point, these two phases are not visually distinguishable (see \f{fig-1}) as discussed above, making the task much harder. The same is true for the Ising critical point\c{Mehta_2019}. Next, by breaking the symmetry between the two different orientations, we induce a first-order phase transition in the system to further show that the classification task gets easier around an abrupt transition. We then train logistic regression and random forest on physical features and compare the results with the models trained on lattice data. Further, we show that one can use simple features rather than the raw snapshots to get better accuracy with simpler ML models. 
 \begin{figure}
    \includegraphics[width=0.82\linewidth]{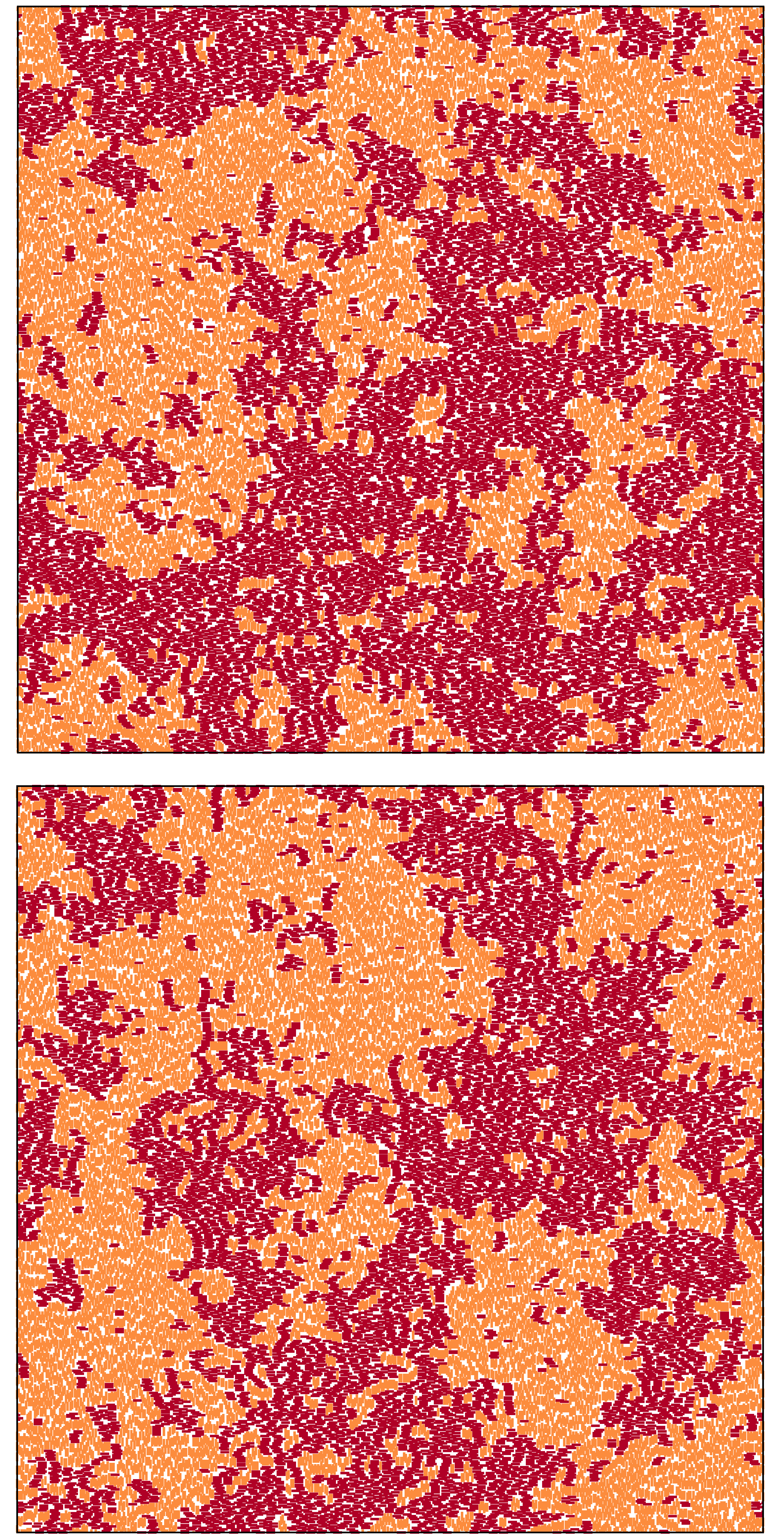}
    \caption[justification=centering]{Typical snapshots of the isotropic phase (top panel) and the nematic phase (bottom panel) close to criticality. Different colors correspond to different orientations.}
    \label{fig-1}
\end{figure}

In the above-mentioned methods of classifying the phases, one must know the critical point in advance to label the data for training. Other methods such as clustering can be used which do not need prior knowledge of the critical point. It is shown that the neural networks can be trained to calculate the critical point with only theoretical ground state snapshots of ordered phase for models like ferromagnetic and anti-ferromagnetic Potts models~\cite{Tan_2020,Tan_2020_2}. Here we employ a similar strategy by training CNN on ordered phase ground state lattice to estimate the critical point.

The rest of the manuscript is organized as follows. We briefly describe the model, the algorithm to generate the equilibrium configurations, and the phenomenology in Sec.~\ref{model}. In Sec.~\ref{second}, we discuss the classification task around the I-N critical point using three machine learning techniques- logistic regression, deep neural networks, and convolutional neural networks that are trained on the I-N transition data. Next, we show the results including the performance of the classification tasks near the second-order and the first-order transitions in Sec.~\ref{res}. We then show how physical features improve the performance of the models like logistic regression and random forests. In Sec.~\ref{critical} we show how the nematic phase information can be used to estimate the critical point of the system of hard rods. Finally, we conclude in Sec.~\ref{conclude}.
\section{Model, Algorithm and the Phenomenology}
\label{model}
 We consider a system of hard rods of length $k$ on a square lattice of size $L\times L$ with periodic boundary conditions. Each rod occupies $k$ consecutive lattice sites along one lattice direction and thus, can have two possible orientations: horizontal and vertical. The only constraint is that no two rods can overlap. All the configurations with no overlap are equally likely. Each rod is associated with an activity $z = e^\mu$ where $\mu$ is chemical potential. The system is treated using grand canonical ensemble where $z$ controls the density (fraction of sites occupied by the rods) of the system. 

 To simulate the system, we use the algorithm presented in this ref.\c{PhysRevE.89.052124}.
 The Monte Carlo algorithm is described below: at each step a row or a column is randomly selected. If a row(column) is selected all the horizontal(vertical) rods in that row are removed. This is the evaporation step. Now we end up with segments of empty sites separated by the forbidden sites due to the vertical rods passing through them. The next step is to reoccupy these empty segments with horizontal rods following correct statistical weights. The problem of occupying the empty row is reduced to filling the empty sites in one dimension. And the probabilities of the new configuration can be  calculated exactly\c{PhysRevE.87.032103,kundu2}. This is a deposition step. A Monte Carlo step consists of $2L$ such evaporation-deposition steps. The equilibration is performed for $6\times 10^5$ MC steps. And then the snapshots of the system are taken at an interval of $5000$ MC steps to ensure they are uncorrelated.
 
\begin{figure}
    \centering
    \includegraphics[width=0.94\linewidth]{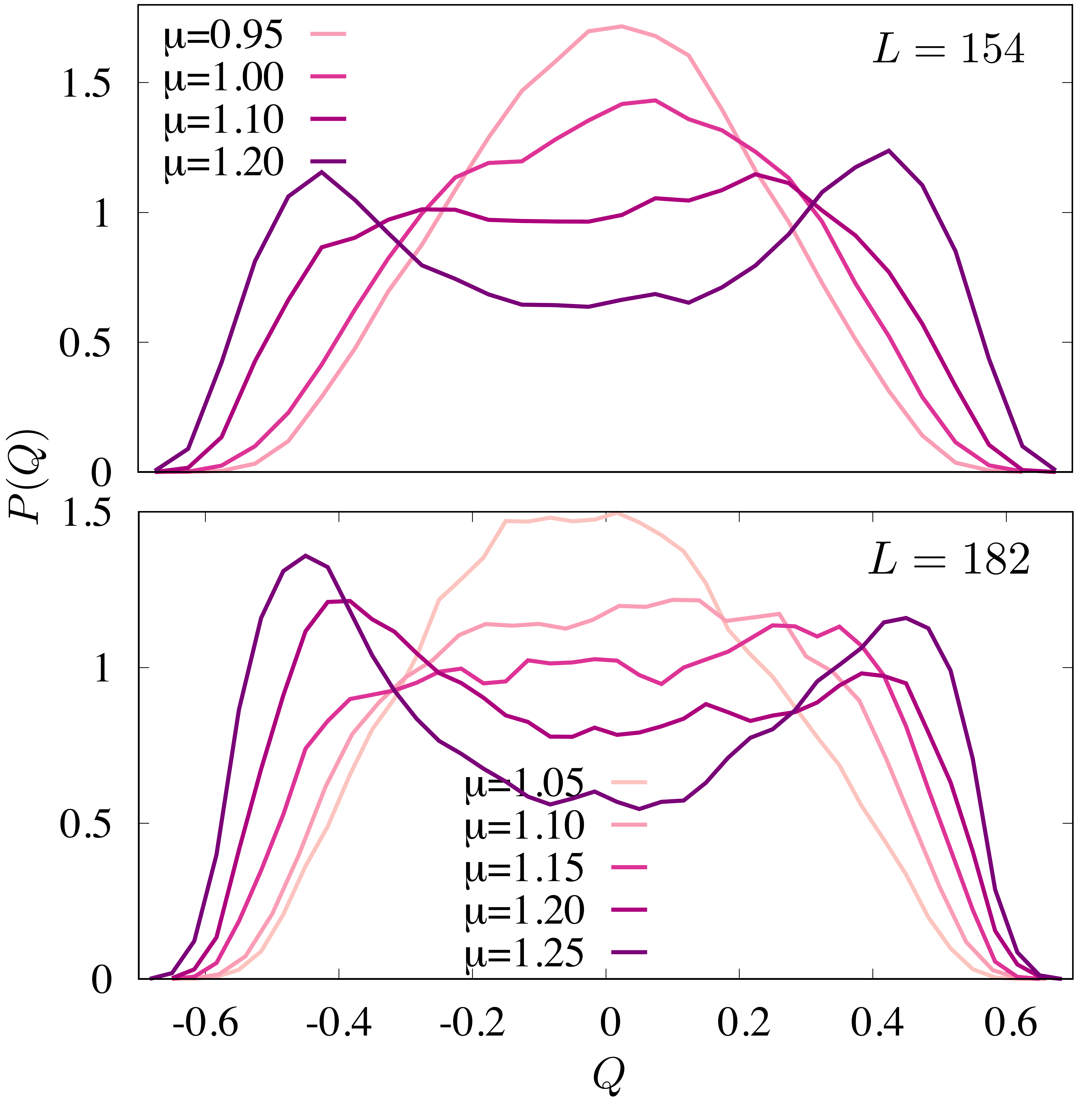}
    \caption[justification=centering]{Probability distribution of order parameter $Q$ at different $\mu$ values around the I-N critical point. The data are for system size $L=154$ (top-panel) and $182$ (bottom-panel).}
    \label{fig-2}
\end{figure}

 The system undergoes two continuous phase transitions with increasing density when $k \geq 7$: first from a low-density isotropic phase to a nematic phase and second, from the nematic phase to a high-density disordered phase. We try to classify the isotropic and the nematic phases around the first transition. The critical chemical potential $\mu_c$ is determined from the probability distribution $P(Q)$ of the order parameter $Q$ defined as $Q=(n_h-n_v)/(n_h+n_v)$, where $n_h$ and $n_v$ are the number of horizontal and vertical rods correspondingly (note for the nematic phase $Q>0$ and for the isotropic phase $Q \approx 0$). Below $\mu_c$, $P(Q)$ has only a single peak around $Q=0$ corresponding to the isotropic phase and above $\mu_c$, $P(Q)$ develops two symmetric peaks corresponding to the nematic phase (see Fig. \ref{fig-2}).
 Further, we study a first-order transition in the system by introducing a variable $\Delta$ (equivalent to the external magnetic field in Ising model) that breaks the symmetry between horizontal and vertical rods. The corresponding chemical potentials for the two types of rods are $\mu_{\rm h}=\mu+\Delta$ and $\mu_{\rm v}=\mu-\Delta$. As $\Delta$ is varied from a negative to a positive value at a given $\mu$, the system undergoes a first-order transition from a horizontal rod rich phase to a vertical rod rich phase (both are nematic). The order parameter $Q$ changes abruptly as shown in \f{fig-3}. In the following sections, we discuss the classification problem around the I-N criticality and the first-order transition point as mentioned above.
 \begin{figure}
    \centering
    \includegraphics[width=0.94\linewidth]{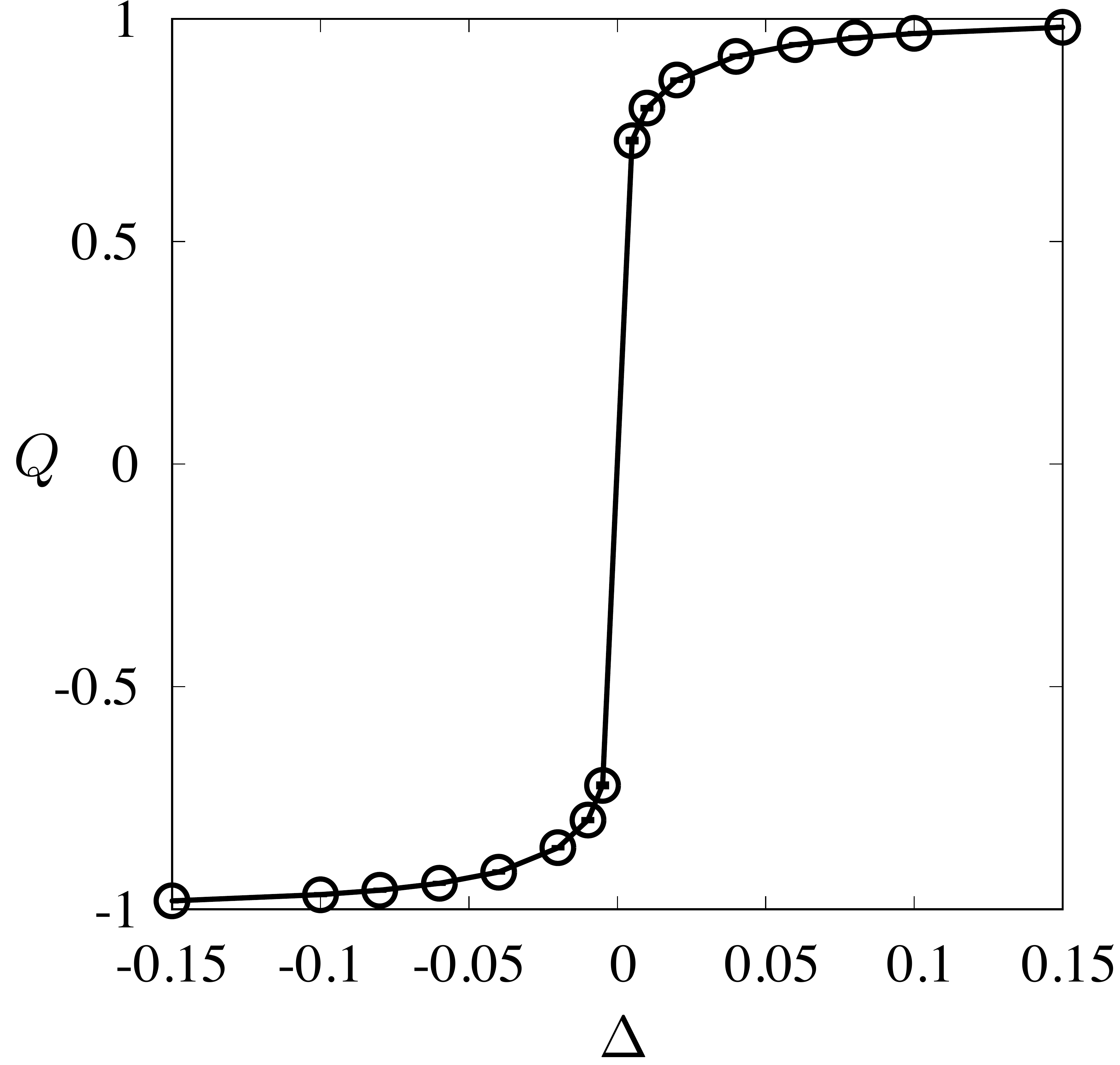}
    \caption[justification=centering]{The variation of the order parameter $Q$ with the field variable $\Delta$ around the first-order transition. The data are for $L=98$.}
    \label{fig-3}
\end{figure}
\section{Classification around the critical point}
\label{second}
Typical snapshots of the system in the disordered and the nematic phases close to the I-N critical point are shown in \f{fig-1}. It is evident from the snapshots that they are not visually distinguishable. Hence, we attempt to classify them using Machine learning. The data around the I-N critical point is trained on logistic regression, deep neural network (DNN) and convolutional neural network (CNN). The data are for lattice size $L = 98$ and rod length $k = 7$. $1500$ snapshots are generated at each $\mu$ value. Equal number of data points are taken on the either side of the critical point $\mu_c = 0.97$ for training. The snapshots below $\mu_c$ as labeled 0 and above $\mu_c$ are labeled 1. The data is divided into 3 parts. First the data is divided into 85\% train set and 15\% test set. The train set is further divided in two sets. The model is trained of 85\% of the train set and validated on 15\% of the train set.

\subsection{Logistic regression}
Logistic regression is the simplest classification algorithm. The algorithm takes the weighted sum of the input features with added bias and evaluates a non-linear sigmoid function. The sigmoid function outputs a number between 0 and 1, representing the probability of the input being phase labeled 1. The loss function is log loss and the optimizer is a stochastic gradient descent optimizer. To train this model the 2D data is flattened to a 1D array and fed into the model. The logistic regression is trained using SGD classifier from Sci-kit library\c{scikit-learn}.
\begin{table}
\begin{tabularx}{0.45\textwidth} { 
  | >{\centering\arraybackslash}
  Y
  | >{\centering\arraybackslash}
  Y
  | >{\centering\arraybackslash}X 
  | >{\centering\arraybackslash}X 
  | >{\centering\arraybackslash}X
  | >{\centering\arraybackslash}X | }
 \hline
 &\multicolumn{3}{|c|}{Accuracy}\\
 \cline{2-4}
 
 $\mu$ & Logistic Regression & DNN & CNN \\
 \hline
0.8245  & 0.545  & 0.620 & 0.880\\ 
\hline
0.8730  & 0.48 & 0.560 & 0.793\\
\hline
1.0670  & 0.435  & 0.555 & 0.796\\
\hline
1.1155  & 0.55  & 0.590 & 0.890\\
\hline
\end{tabularx}
\caption{The table shows the accuracy of different models at four $\mu$ values around $\mu_c = 0.97$ for the system size $L = 98$. Note that the phase is isotropic when $\mu <\mu_c$ and nematic when $\mu > \mu_c$.}
\label{Table-1}
\end{table}

\subsection{Deep Neural Networks}
Deep neural networks (DNNs) have been shown to classify the phases of the Ising model near criticality with accuracy between 0.80 to 0.90\c{Mehta_2019}. At each layer every node takes input from all the nodes from the previous layer. A node calculates the weighted sum of all the inputs with added bias and evaluates a non-linear activation function. These values are fed as input to the next layer. The weights are updated iteratively with an optimizer to decrease the loss function. The DNN we trained consists of 5 layers with 392, 294, 196, 98, 1 nodes in each layer respectively. We chose the number of nodes in each layer to be of the order of the input size. The depth of the network is increased gradually from a single layer until optimal results are obtained. And other hyperparameters are chosen by random search\c{journals/jmlr/BergstraB12}. The flattened 1D data is fed to the network. The Activation function in the first four layers is relu activation and in the last layer is sigmoid activation. The last layer is the same as the logistic regression. The loss function is binary cross-entropy and the optimizer is ADAM. The data is trained on batches of batch size 64. To avoid overfitting $L_2$ regularization is added to each layer. The hyperparameters like the number of layers, number of nodes in each layer, batch size, $L_2$ regularization value are chosen by trial and error method. This Neural network is trained using Keras\c{chollet2015keras} and Tensorflow framework\c{tensorflow2015-whitepaper}.
\subsection{Convolutional Neural Networks}
Convolutional neural networks (CNN) have been shown to classify images and detect objects in the images with very high accuracy\c{Taigman_2014_CVPR}. The main difference between DNN and CNN is the convolution layer. In the convolution layer, a 2D filter is convoluted over the 2D input feature space. The filter slides over the input space and calculates the sum of the element-wise multiplication at each step. Convolution layers are followed by a pooling layer. The pooling layer summarises the feature by averaging (average pooling) or taking the maximum value (max pooling). The filters in the convolution layer detect low-level features at different parts of the input and the convolution layer does not change the spatial structure of the input. The pooling layers decreases the dimensions of the feature space. After a few Convolution $+$ Pooling layers the 2D output is flattened into a 1D array and fed into the fully connected layers. There are many possible combinations of CNNs. The architecture is dependent on the input data and is generally inspired by previously successful networks. Our architecture is inspired from Ref.\c{Sigaki_2020,smith2016deep,alex} and the optimized hyperparameters are chosen after a random search\c{journals/jmlr/BergstraB12}. In this work we chose 2 layers of Convolution + Max pool with $3\times3$ sized 4 filters in each convolution layers and $2\times2$ sized filter with stride $2$ in pooling layers. These are followed by three fully connected layers with 256, 128, 1 nodes respectively. The activation in all layers is relu except for the output layer with sigmoid activation. The loss function is binary cross-entropy, the optimizer is ADAM. To avoid overfitting, $L_2$ regularization term is added to each convolution layer and a dropout layer is added before the fully connected layers. The data is trained in batches. This CNN is trained using Keras\c{chollet2015keras} and Tensorflow framework\c{tensorflow2015-whitepaper}.
\subsection{Random Forest}
Random forest is a simple and powerful machine learning model that can be used for both classification and regression tasks. The main idea of random forest is to build a bunch of decision trees and all these trees as an ensemble will solve the problem at hand. Physical features are given as input instead of lattice snapshot like in earlier mentioned models. Random forest is trained using a random forest classifier from Sci-kit library. Hyper parameters like maximum depth, number of trees are chosen by random search. In this work we chose random forest of maximum depth 4 with 4000 estimators.
\par
\section{Results}
\label{res}
We compare the accuracy of the phase classification problem around the I-N transition for the three models in Table. \ref{Table-1}. It is evident that the CNN outperforms the logistic regression and DNN. This observation is intuitive as CNN is designed to capture the relevant features. Thus, for further study we use CNN with the same architecture and discuss the classification results in detail. We calculate the error bars using Bootstrapping. 
\subsection{Learning near second-order transition}
\label{res_second}
The classification far from the critical point is a trivial task. As we move towards the critical point the classification gets difficult and the performance of the model worsens. This can be seen in Table.  \ref{Table-2}. In \f{fig-4}, we plot the accuracy as a function of the distance from the critical point-- it is evident that the performance of the CNN reduces as we approach the critical point $\mu_c=0.97$. Long wavelength fluctuations and diverging correlation length make the classification task difficult near criticality. 

\begin{table}
\begin{tabularx}{0.45\textwidth} { 
  | >{\centering\arraybackslash}X 
  | >{\centering\arraybackslash}X 
  | }
 \hline
 \multicolumn{2}{|c|}{System size L = 98}\\
 \hline
$\mu$ & Accuracy \\
 \hline
0.50  & 1.000   \\ 
\hline
0.60  & 1.000  \\
\hline
0.70  & 0.985  \\
\hline
0.80  & 0.905  \\
\hline
1.10  & 0.900   \\ 
\hline
1.20  & 0.945  \\
\hline
1.30  & 1.000  \\
\hline
1.40  & 1.000  \\
\hline
\end{tabularx}
\caption{The table shows the accuracy of CNN at different $\mu$ values around the I-N critical point ($\mu_c=0.97$) when $L=98$.}
\label{Table-2}
\end{table}
\begin{figure}
    \includegraphics[width=8cm]{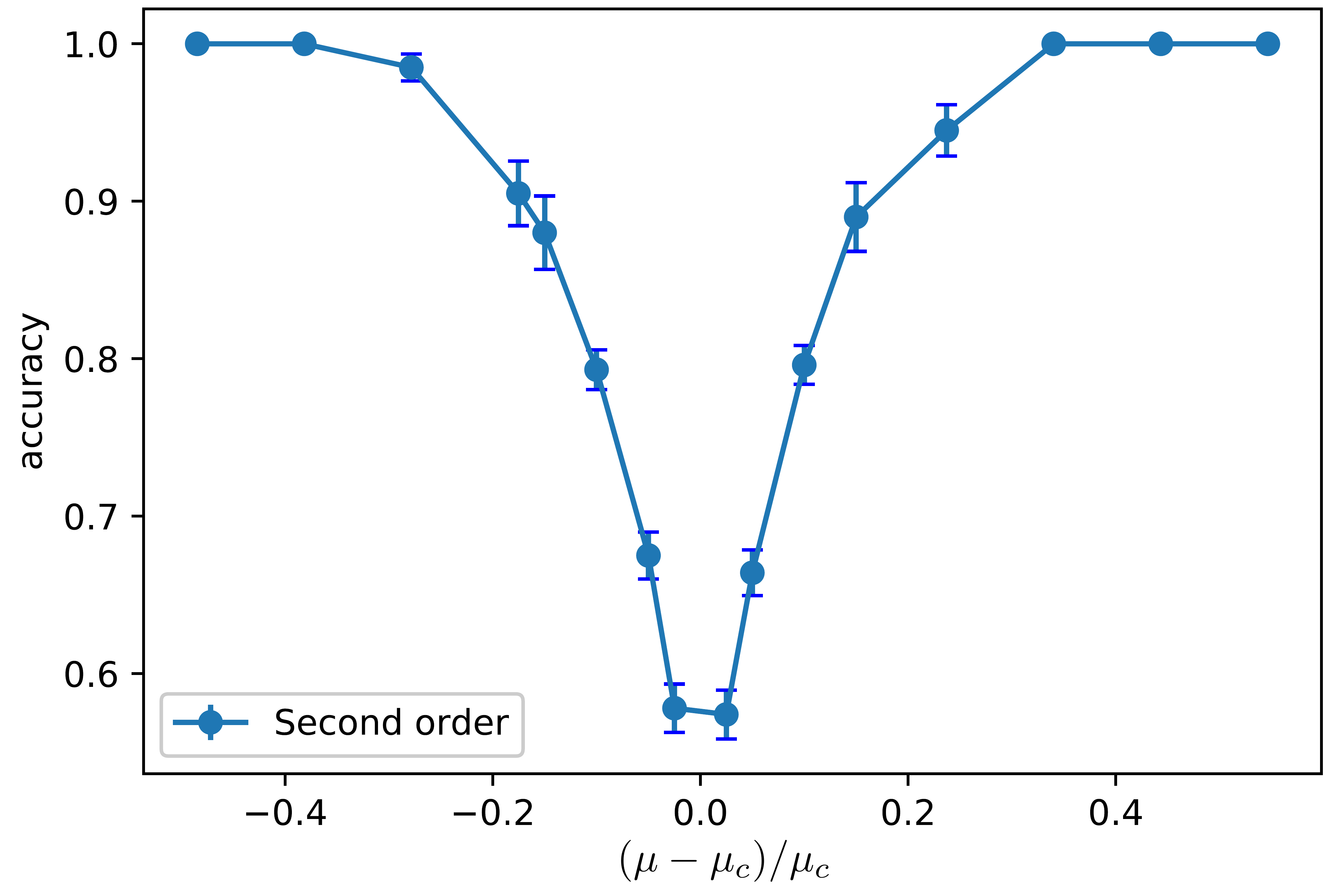}
    \caption{The accuracy of CNN versus the distance of $\mu$ from the critical point $\mu_c$ for system size $L = 98$} 
    \label{fig-4}
\end{figure}
\subsection{Finite size effects}
\label{fs}
 We train the same CNN architecture on larger systems to see how the model performance changes with system size. The additional trained systems sizes are $L = 126, 154$, and $182$. The results in Table. \ref{Table-3} clearly show that as the lattice size increases the model performance also increases. And for a given system size the accuracy decreases as the $\mu$ value approaches $\mu_c$-- see \f{fig-5}. This increase in performance as the system size increase can be attributed to finite-size effects. Close to the critical point the correlation length becomes of the order of the system size. Thus larger system sizes are better in representing the criticality which leads to better predictability of phases compared to the smaller system sizes.
\begin{table}
\begin{tabularx}{0.45\textwidth} { 
  | >{\centering\arraybackslash}Y
  | >{\centering\arraybackslash}X
  | >{\centering\arraybackslash}X 
  | >{\centering\arraybackslash}X
  | >{\centering\arraybackslash}X
  | >{\centering\arraybackslash}X
  | }
  \hline
  &\multicolumn{4}{|c|}{Accuracy}\\

 \cline{2-5}
$\mu$ & $L=98$ & $L=126$ & $L=154$ & $L=182$ \\
 \hline
$0.85~\mu_c$  & 0.880  & 0.945 & 0.956 & 0.985\\ 
\hline
$0.90~\mu_c$  & 0.793  & 0.867 & 0.890 & 0.955\\
\hline
$1.10~\mu_c$  & 0.796  & 0.853 & 0.902 & 0.937\\
\hline
$1.15~\mu_c$  & 0.890  & 0.955 & 0.962 & 0.990\\
\hline
\end{tabularx}
\caption{The table shows the accuracy at four $\mu$ values which are $\pm 10\%$ and $\pm 15\%$ away from corresponding $\mu_c$ for system sizes $L = 98, 126, 154$, and $182$.}
\label{Table-3}
\end{table}
\begin{figure}
    \includegraphics[width=8cm]{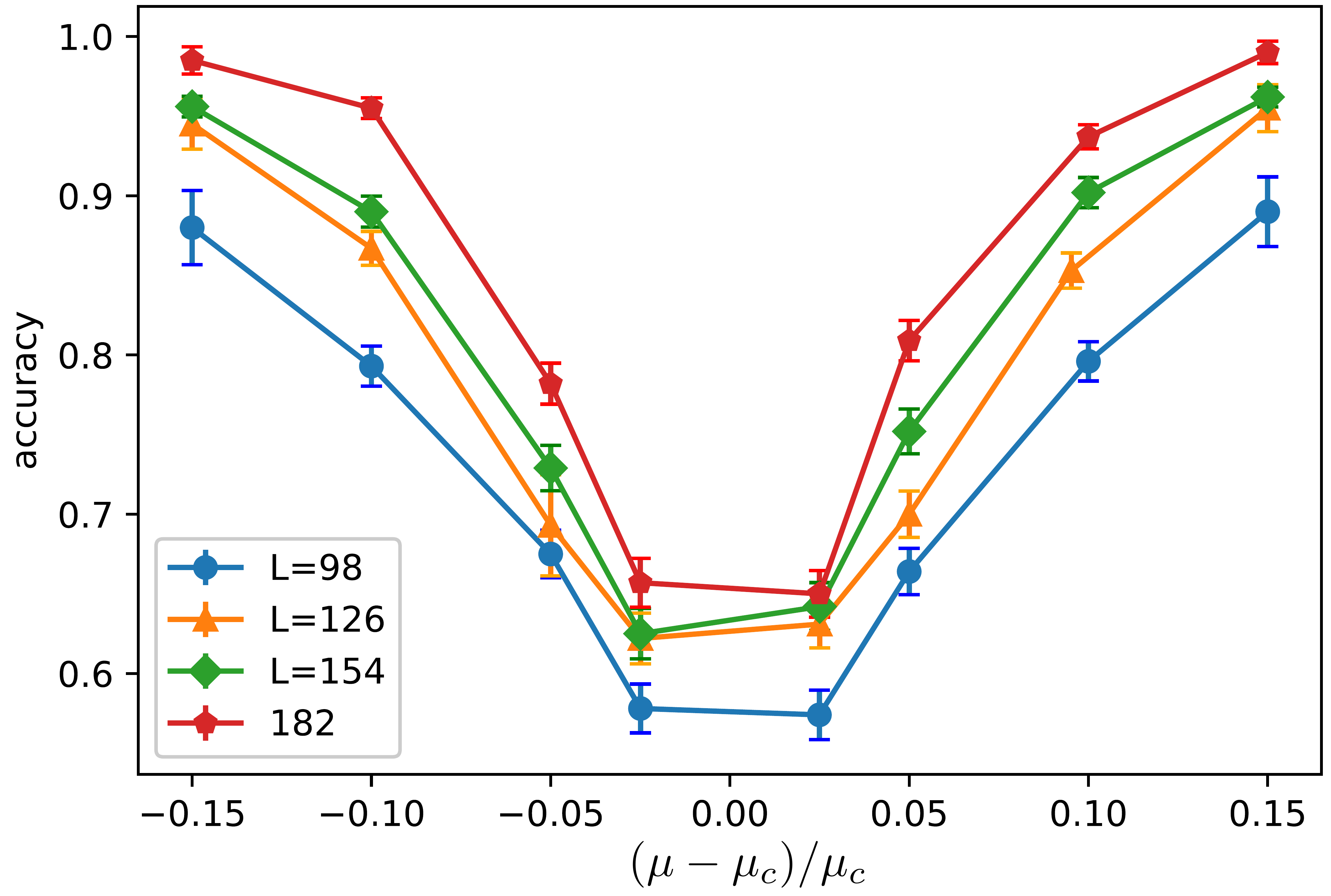}
    \caption{The accuracy of CNN for different system sizes as a function of the distance from the critical point.}
    \label{fig-5}
\end{figure}
\subsection{Learning near First-order transition}
\label{first}
\begin{figure}
    \includegraphics[width=8cm]{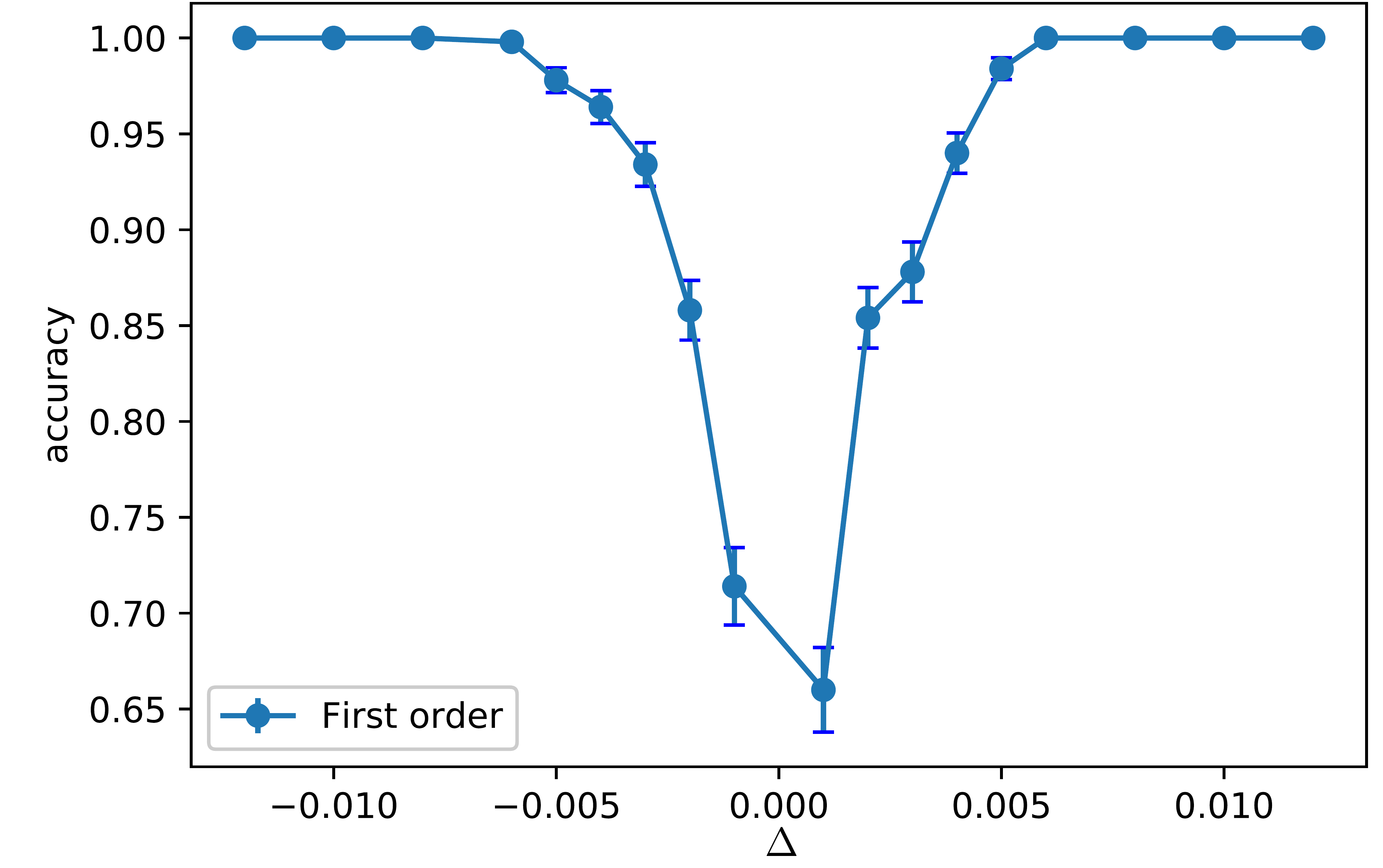}
    \caption{The accuracy of CNN at different values of $\Delta$ for system size $L = 98$. Here we set $\mu=1.02$.}
    \label{fig-6}
\end{figure}

To study the first-order transition we use the variable $\Delta$ as discussed in Sec.~\ref{model}. The data are trained on the same CNN architecture. The order parameter in this case changes abruptly unlike in the case of a second-order transition. Hence, it should be easier for the model to classify the phases around a first-order phase transition. The accuracy of the model as a function of the distance from the transition point is presented in Table. \ref{Table-4} and in Fig.\ref{fig-6}. On comparison, the model (CNN) performs better in classifying phases around the first-order transition (this can be seen from the \f{fig-4} and \f{fig-6})-- in case of the first-order transition, the accuracy reaches $100\%$ when the relative distance from the transition point $|\Delta|/\mu \gtrsim 0.008$, while in case of the second-order transition, the same quantity $|\mu-\mu_c|/\mu_c \gtrsim 0.36$ (for $L=98$). 
Thus, the model is able to classify the phases around the first-order transition at $\mu$ values that are very close to the transition point with very high accuracy.
\begin{table}
\begin{tabularx}{0.45\textwidth} { 
  | >{\centering\arraybackslash}X 
  | >{\centering\arraybackslash}X 
  || >{\centering\arraybackslash}X
  | >{\centering\arraybackslash}X
  | }
\hline
$\Delta $ & Accuracy & $\Delta $ & Accuracy   \\
\hline
-0.001  & 0.714  & 0.001 & 0.660 \\ 
\hline
-0.002  & 0.858   & 0.002 & 0.854 \\
\hline
-0.003  & 0.934  & 0.003 & 0.878 \\
\hline
-0.004  & 0.964  & 0.004 & 0.940 \\
\hline
-0.005 & 0.978  & 0.005 & 0.984 \\
\hline 
-0.006 & 0.998  & 0.006 & 1.000 \\
\hline
-0.008 & 1.000  & 0.008 & 1.000 \\
\hline
-0.010 & 1.000  & 0.010 & 1.000 \\
\hline
-0.012 & 1.000  & 0.012 & 1.000 \\
\hline
\end{tabularx}
\caption{The table shows the accuracy of CNN at different values of $\Delta$ for system size $L = 98$. Here we set $\mu=1.02$.}
\label{Table-4}
\end{table}
\subsection{Classification using physical features}
\begin{figure}
    \includegraphics[width=8cm]{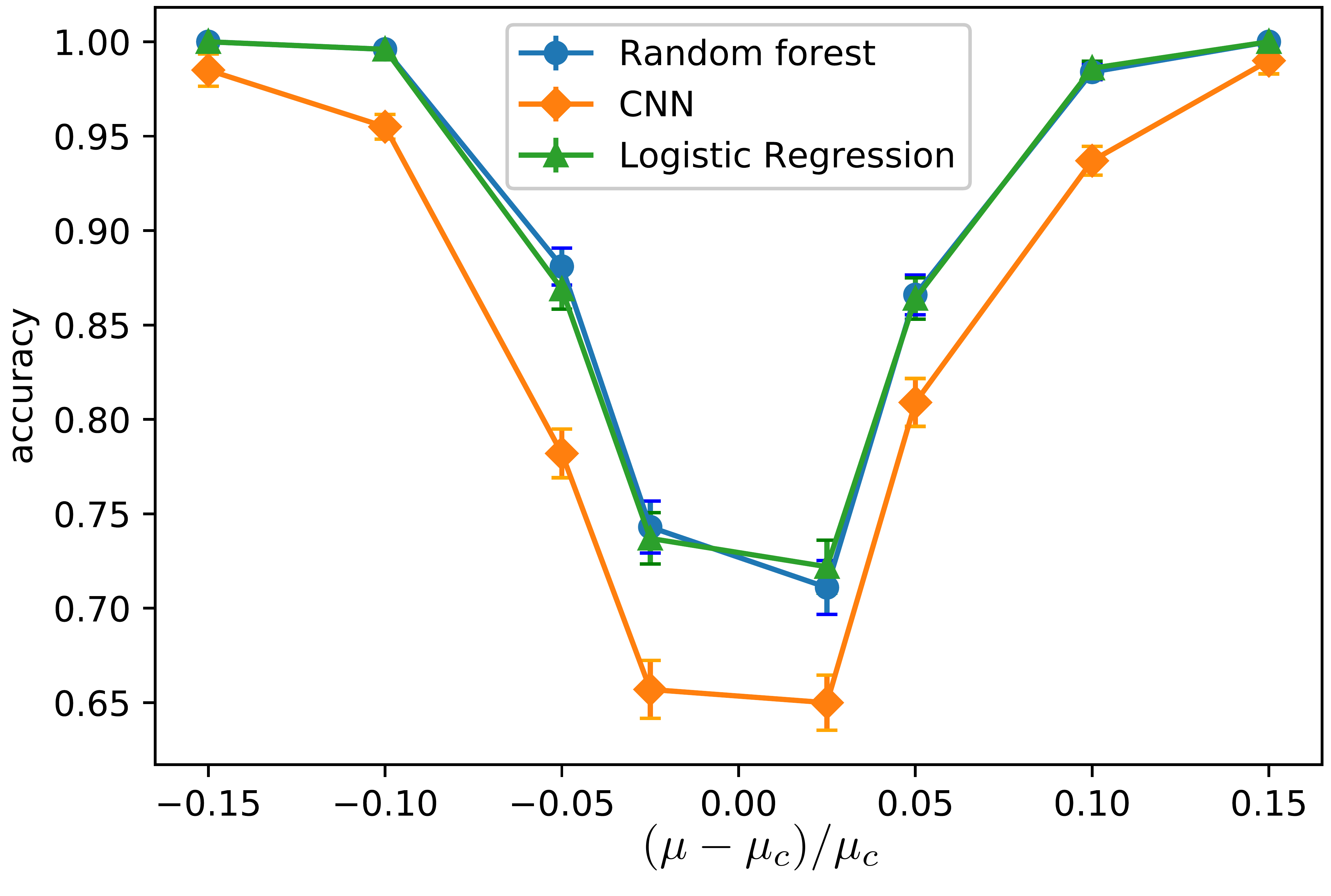}
    \caption{The accuracy of Random forest, Logistic regression (trained using the {\em physics-guided} features) and CNN (trained on snapshots) for the system size $L = 182$.}
    \label{fig-7}
\end{figure}

 In this section, we see how the accuracy of the models changes if some physical features are given as inputs instead of lattice snapshots. The physical features we chose are density (fraction of occupied sites) and order parameter (as defined above). These can be calculated from the snapshots. We train logistic regression and random forests with these two input features. Upon comparing the results in Table.\ref{Table-5} and Table.\ref{Table-3}, it is evident that the performance of both the models are better than that of the CNN trained on the configurations or snapshots. One should note that if we use regular snapshots as inputs, the simpler models like Logistic regression and Random forest perform poorly as discussed before. It is evident from Fig.\ref{fig-7} that Logistic regression and Random forests trained on physical features outperforms the CNN model. This result also infers that these ML models fail to capture the complex correlations that represent criticality. To improve the performance near a critical point one needs to include more complex features distinguishing the two sides of criticality. 
\begin{table}
\begin{tabularx}{0.45\textwidth} { 
  | >{\centering\arraybackslash}X 
  | >{\centering\arraybackslash}X 
  | >{\centering\arraybackslash}X
  | }
\hline
$\mu$ & Logistic regression & Random forests \\
\hline
$0.85~\mu_c$  & 1.00 & 1.00\\ 
\hline
$0.90~\mu_c$  & 0.996 & 0.996\\
\hline
$0.95~\mu_c$  & 0.869 & 0. 881\\
\hline
$0.975~\mu_c$  & 0.737 & 0.743\\
\hline
$1.025~\mu_c$ & 0.722 & 0.711 \\
\hline 
$1.05~\mu_c$ & 0.864 & 0.866 \\
\hline
$1.10~\mu_c$ & 0.986 & 0.984\\
\hline
$1.15~\mu_c$ & 1.00 & 1.00\\
\hline
\end{tabularx}
\caption{The table shows the accuracy of logistic regression and random forest at different $\mu$ values for system size $L = 182$}
\label{Table-5}
\end{table}
\section{Estimating critical point using ML}
\label{critical}
\begin{figure}
    \includegraphics[width=8cm]{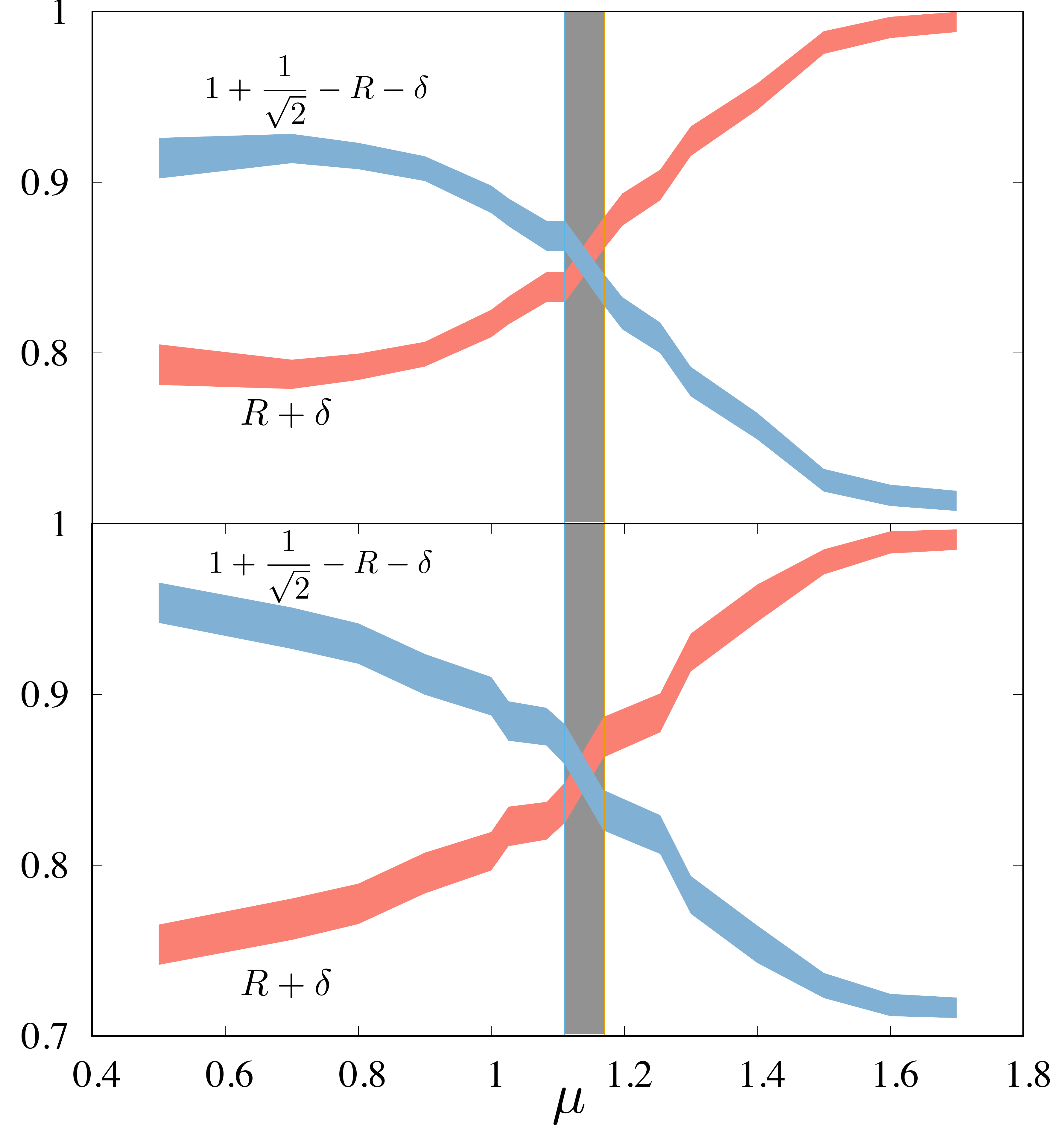}
    \caption{$R+\delta$ and $1+1/\sqrt 2-R-\delta$ as a function of $\mu$ for $L=182$. Top panel: the data are for CNN when snapshots are used as inputs. Bottom panel: the data are for Logistic regression when physics-guided features are used as inputs. The intersection of these two curves coincides with the estimated value of the critical point obtained from the distribution of the order parameter $Q$ as denoted by the vertical thick line. Thickness denotes the corresponding error bar.}
    \label{fig-8}
\end{figure}

To estimate the critical point of the system of hard rods we train neural networks on ground state snapshots of the ordered phase. This method is successfully used to estimate critical points of models like ferromagnetic and anti-ferromagnetic Potts models, 3D classical $O(3)$, 2D XY models~\cite{Tan_2020,Tan_2020_2}. Nematic phase configurations of the system size $182\times182$ are used as the training set. The data are generated as mentioned in Sec.~\ref{model} by breaking the symmetry between horizontal and vertical rods upon introducing the variable $\Delta$. Snapshots are generated at $\mu = 1.50$ and $\Delta = 0.10$. At this value of $\Delta$, the system becomes almost fully ordered ($Q \approx 0.98$). All the vertical (horizontal) rods in the horizontal (vertical) rich snapshots are removed to obtain snapshots in which rods are completely aligned in one direction with order parameter $Q$ being $1$. The vertically aligned snapshots are labeled $[1,~0]$ and the horizontally aligned snapshots are labeled $[0,~1]$. Note that the nematic phase has two possible configurations-- all horizontal and all vertical. CNN is trained using this data with one convolution layer followed by max pooling layer and two dense layers with softmax activation in the output layer, $L_2$ regularization is used to avoid overfitting. The loss, optimizer and activation used in training this model are categorical cross-entropy, adam optimizer, and relu activation. We also train Logistic regression with the physics-guided features as discussed above. In this case we work with $Q$ values $\leq 1$ (almost fully ordered). These trained models are then used to predict the labels of snapshots over a range of $\mu$ values. The norm of the predicted label, $R$, is calculated. The true value of $R$ can vary from 1 to $1/\sqrt 2$ where $1$ corresponds to the fully ordered nematic phase and $1/\sqrt 2$ corresponds to the case when the model fails to classify the phase as horizontal or vertical rich nematic phase. As the model is trained only with (almost) fully ordered phases, if one tests the output for a disordered phase, the output vector would ideally be $[0.5,~0.5]$ and the value of $R$ would be $1/\sqrt 2$. The norm of the predicted labels of ordered states which are fully packed with rods in one direction should ideally be 1. To correct this, the difference between $1$ and the $R$ value of the predicted labels of the fully packed nematic phase, $\delta$ [where $\delta = 1-(R_{\rm ver}+R_{\rm hor})/2$ and $R_{\rm ver}$($R_{\rm hor}$) is the norm of the predicted label of the fully packed nematic phase snapshot corresponding to vertical (horizontal) rods] is added to $R$. This $R$ is plotted against $\mu$ values to estimate the critical point. Assuming linearity of $R$ with $\mu$ near criticality, $\mu_c$ should be associated with the mid-point and is given by the intersection of the two curves $R + \delta$ and $1+1/\sqrt2 - R - \delta$. As shown in the Fig \ref{fig-8}, the intersection point of $R + \delta$ and $1+1/\sqrt2 - R - \delta$ is the estimated critical point. The estimated values of $\mu_c$ by both the models are in close agreement (within error bars) with the numerical value $\mu_c = 1.14 \pm 0.03$ obtained from the probability distribution of the order parameter (see Fig.~\ref{fig-2})

In addition, the modulus of the difference between two elements of the output vector $R$, can be treated as a {\em Machine learned} order parameter-- the ordered phase will correspond to a value $\approx 1$, and in the isotropic phase it will be $\approx 0$. 
\section{Conclusion}
\label{conclude}
In this work, we classify the phases of the system of hard rods on a square lattice. Although the classification task is trivial far from the transition point,  sufficiently close to criticality system spanning fluctuations set in, making the phases visually indistinguishable and thus, the classification problem becomes harder. Three machine learning models, logistic regression, deep neural network, and convolutional neural network are trained to classify the phases around the isotropic-nematic transition. CNN has been shown to classify the phases with higher  accuracy than the other two methods. We showed that the model performance improves with the increase in the system size-- this is attributed to the finite-size effects. We further induce a first-order phase transition into the system using a field variable and the CNN is trained to classify phases around the transition. We demonstrate that classifying phases is easier around a first-order transition than around a second-order transition. Although the classification problem around the first-order transition is not the same as that around the second-order transition, our conclusions are not affected by that. We have also shown that physics-guided features drastically improve the performance of simpler models like logistic regression and random forest. In fact, with such feature engineering, they outperform more complex models like CNN (where the raw snapshots are used as inputs). We then estimate the critical point of the system of hard rods $\mu_c$ using only the information about the ordered phase. This estimate is in strong agreement with the value calculated by traditional methods. As discussed above, the {\em Machine learned} order parameter can be used to perform the finite size scaling analysis to compute the critical exponents. Our work infers that these methods may further be useful in studying transitions in more complex systems where defining an order parameter is nontrivial {cite QCD}. Another interesting question that emerges from our work is how to capture the critical correlations near a continuous phase transition using ML techniques. These will be future areas of investigation.
\begin{acknowledgments}
We thank Sudhir N. Pathak for useful discussions. JK acknowledges support from the IIT Hyderabad Seed grant. 
\end{acknowledgments}
%

\end{document}